\documentclass[aps,prl,twocolumn,superscriptaddress]{revtex4-1}
\usepackage{graphicx}
\usepackage{epstopdf}
\usepackage{color}
\usepackage{bm}
\usepackage{amsmath}
\usepackage{changes}
\usepackage[colorlinks=true,linkcolor=blue,urlcolor=blue,citecolor=blue]{hyperref}

\definecolor{pink}{rgb}{1,1,0} 
\definecolor{red}{rgb}{1,0,0}
\definecolor{yellow}{rgb}{1,1,0}
\definecolor{orange}{rgb}{1,0.5,0}
\definecolor{green}{rgb}{0,1,0}
\definecolor{blue}{rgb}{0,0,1}
\definecolor{white}{rgb}{1,1,1}
\definecolor{purple}{rgb}{0.5,0,0.5}

\begin{document}
\title{Magnetoelectric effect in dipolar clusters}
\author{Paula Mellado}
\affiliation{School of Engineering and Sciences, 
	Universidad Adolfo Ib{\'a}{\~n}ez,
	Santiago, Chile
}
\author{Andres Concha}
\affiliation{School of Engineering and Sciences, 
	Universidad Adolfo Ib{\'a}{\~n}ez,
	Santiago, Chile
}

\author{Sergio Rica}
\affiliation{School of Engineering and Sciences, 
	Universidad Adolfo Ib{\'a}{\~n}ez,
	Santiago, Chile
}

\begin{abstract} 
We combine the anisotropy of magnetic interactions and the point symmetry of finite solids in the study of dipolar clusters as new basic units for multiferroics metamaterials. The hamiltonian of magnetic dipoles with an easy axis at the vertices of polygons and polyhedra, maps exactly into a hamiltonian with symmetric and antisymmetric exchange couplings. The last one gives rise to a Dzyaloshinskii-Moriya contribution responsible for the magnetic modes of the systems and their symmetry groups, which coincide with those of a particle in a crystal field with spin-orbit interaction. We find that the clusters carry spin current and that they manifest the magnetoelectric effect. We expect our results to pave the way for the rational design of magnetoelectric devices at room temperature.

\end{abstract}

\maketitle
\emph{Introduction.} 
The conciliation of crystal symmetry and magnetic phenomena has been a key element in the understanding of matter and the quest for new materials.  Magnetic degrees of freedom coupled to physical symmetry preclude the realization of magnetic phases, that may manifest as ferromagnetic, ferrimagnetic, and superconductor materials among others \cite{savary2016,rosa2017,edwards1975}. The magnetic structures, correlated in their own right, respond to external fields through a variety of fashions \cite{mellado2012} that may arise piezomagnetism, pyromagnetism and magnetolectricity. The magnetoelectric (ME) effect is especially intriguing \cite{dzyaloshinskii1960,tachiki1961} because the magnetic field ($\bm{H}$) controls the electric polarization ($\bm{P}$) and the electric field ($\bm{E}$) controls the magnetization ($\bm{M}$), through magnetic modes that break both space-inversion ($\mathcal{I}$) and time-reversal ($\mathcal{T}$) symmetries \cite{landauelectrodynamics, spaldin2008,katsura2005}. 

ME effect was first observed in antiferromagnetic chromium oxide, $\rm{Cr_2O_3}$ \cite{astrov1960, astrov1961}. As for $\rm{Cr_2O_3}$ and other ME materials, an applied magnetic field induces not only magnetization, but also electric polarization \cite{dzyaloshinskii1958}. In the linear ME effect, the induced $\bm{M}$, an axial first rank tensor, is linearly proportional to the applied $\bm{E}$, a polar first rank tensor, through the ME coefficients which are matrix elements of the ME axial tensor, $\underline{Q}$ \cite{siratori1992}. The form of $\underline{Q}$ is determined by the transformation of spins on a given lattice under the symmetry operations of the respective crystallographic space-group. Up to date, several magnetic materials have been reported to realize ferroelectricity induced either by spiral magnetic orders or by other modulated or chiral spin arrangements like conical or screw spin structures that break inversion symmetry \cite{spaldin2008,kimura2005}.  Examples include orthorhombic perovskite manganites, hexagonal hexaferrites and cuprates to name a few \cite{kimura2012,matsubara2015}. Nevertheless, in most, the ME effect is hard to detect as it is hidden by magnetic disorder or  other competing phenomena, it manifests at very low temperatures or it is too weak becoming negligible in the view of stronger effects \cite{kimura2007,kitagawa2010,mostovoy2006}. 

\begin{figure}
\includegraphics[width=\columnwidth]{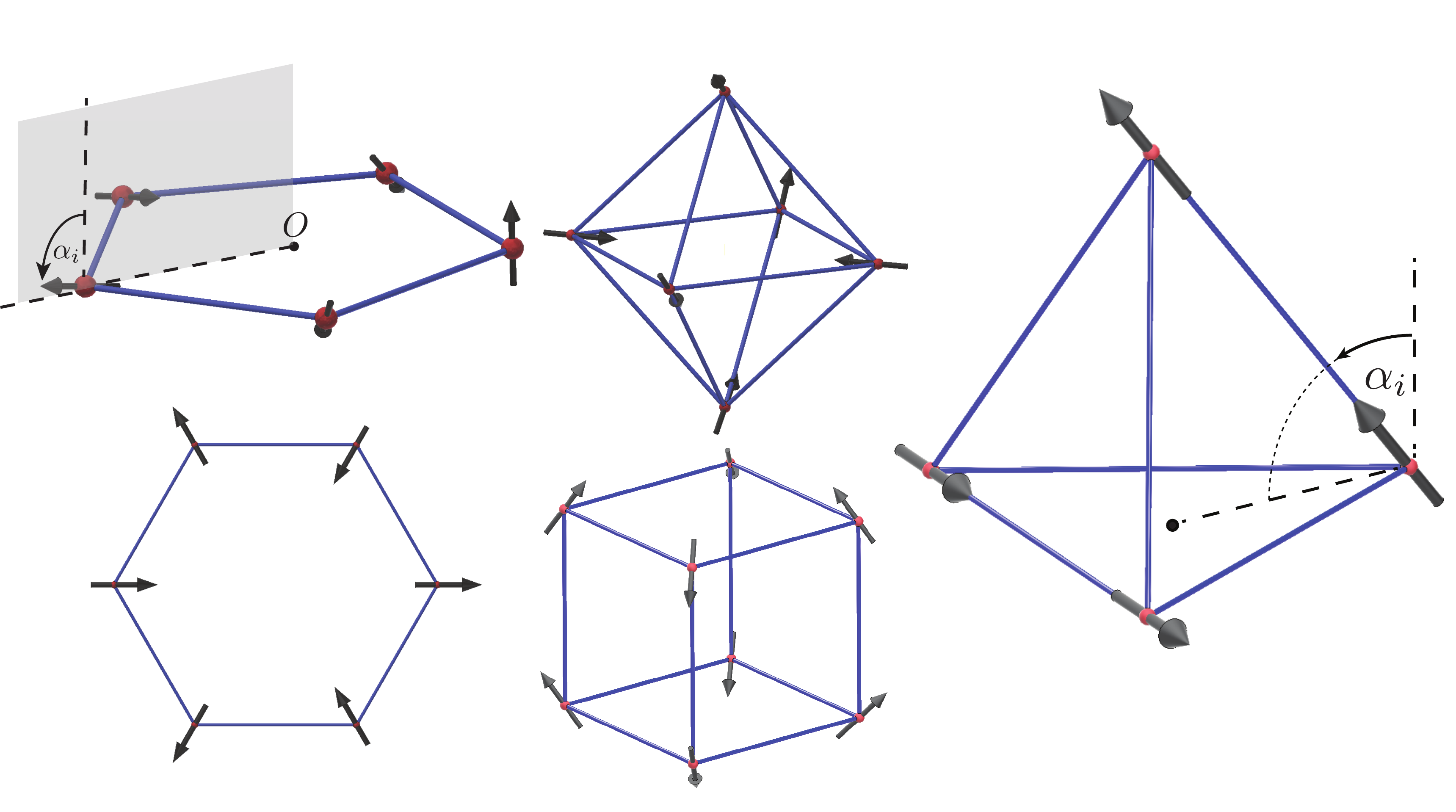}
\caption{Minimum energy magnetic configurations of pentagonal, hexagonal, tetrahedral, cubic and octahedral dipolar clusters. The angle of rotation $\alpha$ and the easy plane of rotation are shown.}
\label{fig:f1}
\end{figure}

\emph{Main results.} 
In this article, we find new simple systems/mechanisms that may realize the ME effect at room temperature through magnetic modes induced by the interplay of dipolar interactions and geometrical constraints. More precisely, we study the ME effect in dipolar systems realized by regular polygons and polyhedra decorated with easy axis magnetic dipoles at their vertices.  The dipolar hamiltonian is mapped exactly into a symmetric and antisymmetric contribution, where the antisymmetric part takes the form of a \textit{Dzyaloshinskii-Moriya} (DM) interaction. Energy minimization of the dipolar energy yields the lowest energy magnetic configurations of regular $n$-sided polygons and several platonic solids. We demonstrate that the magnetic states realize the ME effect and possess multipolar moments, spin current and ME polarization.  Exact diagonalization of the interaction matrix of the ground state sectors yields double degenerate spectra. These degeneracies do not match the dimensions of the pertinent point groups. For example, in the tetrahedral cluster with lowest magnetic energy mode shown in Fig.~\ref{fig:f1}, the  dimensions of the irreducible representations (\textit{irreps})  of the tetrahedral point group, {\it{23}}(T) (hereafter we use international notation for point groups, and include Schoenflies notation in parenthesis) are 1,1,1 and 3 \cite{dresselhaus2007}. The  two non-trivial subgroups of \textit{23} compatible with the magnetic configuration, have only one dimensional \textit{irreps} each. 
The degeneracy of the  eigenvalues are tied to the symmetry and not to the specific form of the hamiltonian, thus we have map the ground state (GS) sector in each case, into an effective hamiltonian $\mathcal{\hat{H}}_{f}$ where collinear dipoles are coupled via Ising-like interactions.   The spectrum of $\mathcal{\hat{H}}_{f}$ yields doublets and more important, $\mathcal{\hat{H}}_{f}$  reflects the symmetries of the GS of the dipolar clusters. We found that the symmetry group of regular polygons with several vertices $n = 2(2s+1)$ ($s>0$ and integer) is the double chiral dihedral point group \textit{n22}($\tilde{D}_{n}$). Its generators allowed to determine $\underline{Q}$ that in this case is diagonal, symmetric and has two independent coefficients.  For regular polygons with $n=4s$, the symmetry group corresponds to the double point group \textit{$\bar{n}$2m}($D_{\frac{n}{2}d}$) which yields symmetric diagonal $\underline{Q}$ with one independent coefficient.
The outcome for regular polyhedra is related to regular polygons. Indeed the  symmetry group of the cube and the octahedron is the double group \textit{$\bar{4}$2m}($\tilde{D}_{2d}$) and thus in both cases $\underline{Q}$ is diagonal with a single independent coefficient.  For the tetrahedron the symmetry group is the double group \textit{422}($\tilde{D}_4$) and $\underline{Q}$ has two independent matrix elements along the diagonal.  
Double groups are subset of $\rm{SU(2)}$ and arise in systems with half-integer  angular momentum and spin orbit interaction. 

 \emph{Magnetoelectric effect.}  
The ME effect can be introduced via an expansion of the free energy in terms of $\bm{H}$ and $\bm{ E}$ \cite{siratori1992}, namely
$\mathcal{F}(\bm{E},\bm{H})=F_0-\frac{\epsilon_{ij}}{8\pi} E_iE_j-\frac{\mu_{ij}}{8\pi} H_iH_j-Q_{ij}E_iH_j+\dots$
where $\epsilon_{ij}$, $\mu_{ij}$ are, respectively, the dielectric
and the magnetic permeability. Derivative of $\mathcal{F}$ in $\bm{H}$ gives $\bm{M}$ and derivative of $\mathcal{F}$ in $\bm{E}$ gives $\bm{P}$, therefore in the linear ME effect $\bm{P}=\underline{Q}\bm{H}$ and $\bm{M}=\underline{Q}\bm{E}$ in proper units. The ME tensor changes sign upon $\bm{r} \to - \bm{r}$ or $t\to -t$, so that a linear ME requires a simultaneous violation of  $\mathcal{I}$ and $\mathcal{T}$ symmetries. To describe ME effect in terms of observable order parameters a common approach is to associate the shape of $\underline{Q}$ to ME moments that arise from the magnetic multipolar expansion \cite{spaldin2008}. 

Expanding the magnetization energy in an inhomogeneous magnetic field, $\bm{H}$,  in powers of the field gradients at some reference point:  $\mathcal{H}_{int}=-\int( \bm{m} ({\bm r}) \cdot\bm{H}(0)-x_i m_j\partial_iH_j(0)) \, d^3\bm{r}-\dots =-\bm{M}\cdot\bm{H}(0)-a(\bm{\nabla}\cdot \bm{H})|_{ \bm{r} =0}-\bm{t}\cdot(\bm{\nabla}\times \bm{H})|_{{\bm r}=0}-q_{ij}(\partial_iH_j+\partial_j H_i)|_{\bm{r} =0}-\dots$, one identifies directly~:
 $a=\frac{1}{3}\int \bm{r} \cdot \bm{m}({\bm r}) \,d^3 \bm{r}$ as a monopolar moment; 
${\bm t}=\frac{1}{2}\int \bm{r} \times \bm{m}({\bm r})\,d^3 \bm{r}$, as a toroidal moment dual to the antisymmetric part of the tensor $\partial_iH_j$; and, a traceless symmetric tensor $q_{ij}=\frac{1}{2}\int (x_i m_{j}+x_j m_i-\frac{2}{3}\delta_{ij}\bm{r}\cdot\bm{m(r)})d^3 \bm{r}$ that describes the quadrupole magnetic moment of the system.

A microscopic mechanism connecting the electric dipole with the spin operator is the spin-orbit interaction that transfers anisotropy from the real space into the spin space. The ME effect and the spin current $\bm{\mathcal{J}}_s\propto \bm{s_i}\times \bm{s_j}$ are directly related in non-collinear spin structures as for instance the spiral state.  In magnets, $\bm{\mathcal{J}}_s$ is associated with the spin rigidity and it is induced between two spins with generic non-parallel configurations. In Ref. \cite{katsura2005} it has been shown that $\bm{\mathcal{J}}_s$ in noncollinear magnets leads to the electric polarization $\bm{\mathcal{P}}\propto \bm{e}_{ij}\times \bm{\mathcal{J}}_s$, where $\bm{e}_{ij}$ is the director vector joining spins $\bm{s_i}$ and $\bm{s_j}$. We show the classical correspondence of the spin current in the supplemental information \cite{supp}.

\emph{The model.}  
The dipolar  \emph{classical} hamiltonian for the systems, in units of Joule $\rm[J]$, reads 
\begin{eqnarray}
\mathcal{H}_{dip}=\frac{\gamma}{2} \sum_{i\neq k=1}^n \frac{\hat {\bm  m}_i \cdot\hat {\bm m}_k - 
3 (\hat {\bm m}_i \cdot \hat {\bm{e}}_{ik} )(\hat {\bm m}_k\cdot \hat {\bm{e}}_{ik} )}{|{\bm r}_i -{\bm r}_k |^3},
\label{eq:Energy}
\end{eqnarray}
 here $\hat {\bm{e}}_{ik}= ({\bm r}_i -{\bm r}_k ) /|{\bm r}_i -{\bm r}_k |$, and $\gamma =\frac{\mu_0 m_0^2}{4\pi}$ has units of $[\rm Nm^4]$ and contains the physical parameters involved in the energy such that $\mu_0$, the magnetic permeability, and ${m_0}$, the intensity of the magnetic moments with units $\rm[m^2 A]$. From now on we normalize all distances by the cluster side length $L$, that is  $\hat {\bm x}_i={\bm{r}_i}/{L}$. Dipoles magnetic moments are normalized by  ${\bm m}_i = m_0   \hat{{\bm m}}_i $, have unit vector:
$ \hat{{\bm m}}_i  = ( \sin\alpha_i \cos\varphi_i ,\sin\alpha_i\sin\varphi_i ,\cos\alpha_i)$, and are located at the vertices $\hat {\bm x}_i$ of regular polygons or platonic solids. They rotate in an easy plane described in terms of a polar angle $\alpha_i$ chosen respect to the $\hat{z}$ axis, and a fixed azimuthal angle $\varphi_i$ that accounts for the projection in the $\hat{x}-\hat{y}$ plane of the vector joining the site $i$ with the centroid of the cluster. 

It is straightforward to show that the dipolar energy of our dipolar  clusters is separable into symmetric and antisymmetric exchange contributions. Indeed for odd polygons,
\begin{eqnarray}
E^{\rm (odd)}&=&\gamma \sum_{ k=1}^{s}  \frac{1}{\Delta_{k}^3}  \Big[\sum_{i=-s}^s \hat {\bm m}_i \cdot\hat {\bm m}_{i+k} + \nonumber \\&&  + \,
 \frac{3}{2} \tan\left(\frac{\pi}{n} k\right)\sum_{i=-s}^s\left(\hat{\bm m}_i\times\hat {\bm m}_{i+k} \right)\cdot \hat {\bm z}\Big] ,
\label{eq:EnergyOdd}
\end{eqnarray}
with $\Delta_k =  \sin \left( \frac{\pi}{n} k\right) /{ \sin(\pi/n)}$ and  $s$ is related to the number of vertices via $n= 2 s +1$. For polygons with even number of vertices the dipolar energy reads,
\begin{eqnarray} 
E^{\rm (even)}&=& \gamma\sum_{ k=1}^{ n/2-1}\Big[ \frac{1}{\Delta_{k}^3} \sum_{i=1}^n \hat {\bm m}_i \cdot\hat {\bm m}_{i+k} +\nonumber \\  
& &\frac{3}{2\Delta_{k}^3} \tan\left( \frac{\pi}{n} k\right)\sum_{i=1}^n \left(\hat {\bm m}_i \times\hat {\bm m}_{i+k} \right)\cdot \hat {\bm z} \Big]+\label{eq:EnergyEven} \\
& &\frac{\gamma}{\Delta_{n/2}^3}   \sum_{i=1}^{n/2}   \left[ - 2 \hat {\bm m}_i \cdot\hat {\bm m}_{i+n/2} +3\hat { m}^z_i{ \hat m}^z_{i+n/2} \right].
\nonumber
\end{eqnarray}
The first term in Eq.~(\ref{eq:EnergyOdd}) and Eq.~(\ref{eq:EnergyEven}) is a symmetric exchange interaction between all dipoles. The second term is an antisymmetric exchange, $h_{\rm DM}=\bm{J}_{DM}\cdot\left(\hat{\bm m}_i \times \hat{\bm m}_{i+k} \right)$, with $\bm{J}_{DM}=\frac{3\mu_0}{8 \pi L^3\Delta_{k}^3} \tan\left( \frac{\pi}{n} k\right)\hat {\bm z}$ (in units of $[\frac{\rm N}{\rm A^2 m^3}]$) also known as the Dzyaloshinskii–Moriya interaction \cite{moriya1960}. 
Dipolar energy for even polygons has two additional exchange contributions between dipoles located at opposite vertices in the cluster. These terms are written separately from the main sum because in the limit: $\lim_{k\to n/2}  \tan\left( \frac{\pi}{n} k\right)\to\infty$, together with $ \left(\hat {\bm m}_i \times\hat {\bm m}_{i+k} \right) \to 0$. More important, they compel for opposite dipoles to be in a collinear configuration. The spin orbit interaction shown here,  is also manifested in the energy of polyhedral clusters, as we show in the supplemental information \cite{supp}.

\emph{Classical ground states.} 
For an even regular polygon the ground state can be computed directly from Eq.~(\ref{eq:EnergyEven}) and it is $\alpha_k = (-1)^{k+1} \frac{\pi}{2}$. Fig.~\ref{fig:f1} shows the resulting antiferromagnetic mode for the hexagonal cluster.
For odd polygons, the ground state configuration of Eq.~(\ref{eq:EnergyOdd}) is satisfied by polar angles $(\alpha_{-s},\alpha_{-s+1},...\alpha_{-1},0,\alpha_{1}...\alpha_{s-1},\alpha_{s})$, that satisfy  $\alpha_{-k} = -\alpha_k$ by symmetry and may be computed numerically (See supplemental information \cite{supp}). 

Moreover we considered the dipolar hamiltonian of three platonic solids: the tetrahedron, the cube, and the octahedron. Energy minimization of Eq.~(\ref{eq:Energy}) resulted in the lowest energy magnetic configurations shown in Fig.~\ref{fig:f1}. For the tetrahedron the GS polar angles for all dipoles yield $\alpha^{(t)}_k= \pi/2$. The cube is such that $\alpha^{(c)}_k = (-1)^{k+1}\arctan(1/\sqrt{2})$.  In the octahedron, collinear dipoles have equal $\alpha$ and at all faces the sum of $\alpha$ yields $\pi$. For all polygons and polyhedra the net magnetization along $\hat{z}$, $m_z=\sum_i\cos(\alpha_i)=0$.  In general, point symmetry operations when applied to the polygons and polyhedra studied here may alter the magnetic state beyond the reversal of the orientation of the dipoles in the lowest energy magnetic configurations shown in Fig.~\ref{fig:f1}. In those cases, $\mathcal{T}$ is not enough for restoring the original magnetic configuration. Instead in this paper we apply a projective symmetry analysis to accomplish that goal.

\emph{ Magnetoelectric moments.} 
The moments of finite clusters plays a crucial role toward the implementation of ME effect in two and three dimensional natural or tailored made lattices \cite{delaney2009}. The GS of the dipolar clusters studied here are odd under $\mathcal{I}$ and $\mathcal{T}$, and therefore a non zero ME response is expected. The ME responses for all clusters are summarized in Table~\ref{table:1} and Table~\ref{table:2}. In Table~\ref{table:1} the first three columns show ME moments  $\bm{t}= \sum_{k} \hat{\bm x}_k\times\hat{\bm{m}}_k$,  $ q_{\alpha\beta} = \sum_{k}(\hat x_k^\alpha m_k^\beta+\hat x_k^\beta m_k^\alpha-\frac{2}{3}\delta_{\alpha\beta}\hat{ \bm x}_k\cdot\hat{\bm{m}}_k)$ and $a= \sum_{k}\hat {\bm x}_k\cdot\hat{\bm{m}}_k$. Fourth and fifth columns show the spin current $\bm{\mathcal{J}}_s=\sum_{k=1}^{\frac{n}{2}-1}\bm{j}_s^{(k)}$  with ${\bm j}_s^{(k)}=\hat{\bm{m}}_k\times \hat{\bm{m}}_{k+1}$  \cite{shi2006} and ME polarization $\bm{\mathcal{P}}=\sum_{k=1}^{\frac{n}{2}-1}\hat {\bm{x}}_{k,k+1}\times \bm{j}_s^{(k)}$, where $a$, $\bm t$, and $q$ are given in units of [$m_0 L$], $\bm{\mathcal{J}}_s$  in units of  ${\gamma}/{L^2}$ = [Joule$\cdot$m]  and $\bm{\mathcal{P}}$ in units of ${\gamma}/{L}$ (numerical values can be found in the supplemental information \cite{supp}).
We found that most polygons with $n=2s+1$ vertices realize a toroidal moment in the $x-y$ plane and quadrupolar moment. $\bm{\mathcal{J}}_s$ and $\bm{\mathcal{P}}$ are also manifested in all odd polygons.  
Even polygons have spin current along the $\hat{z}$ axis, $\mathcal{J}_{z}=(n-1) \sin \left(\frac{2 \pi }{n}\right)$ and polarization vector $\bm{\mathcal{P}}= \sin \left(\frac{2\pi }{n}\right)\left(-\cos\left(\frac{\pi }{n}\right)  , \sin\left(\frac{\pi }{n}\right),0\right)$ in the $x-y$ plane. The square has quadrupolar moment $q_{x^2-y^2}=4 \sqrt{2}$ (in units of $m_0 L$) but aside from it,  $q_{\alpha \beta}$, $a$ and $t$ cancel out in all even polygons. The ME response in these cases is not due to multipole moments from the second order terms in the series expansion of $\mathcal{H}_{int}$. Indeed, the antiferromagnetic ground state configuration of hexagonal and octagonal clusters  resemble a magnetic hexapole and octupole respectively.  \\
In three dimensional clusters, there is not eulerian trail, and therefore $\bm{\mathcal{J}}_s$ and $\bm{\mathcal{P}}$ are computed using the hamiltonian path, a trail that visits each site once.  Polyhedra,  except the cube, have toroidal moment along the $\hat{y}$ axis, like the odd polygons. The tetrahedral cluster has $t_y=-1$, spin current and ME polarization along the $\hat{z}$ and $\hat{x}$ axes respectively.  The cube has no moments but $\bm{\mathcal{J}}_s$ and  $\bm{\mathcal{P}}$ in the $x-y$ plane. The octahedral cluster has monopolar, quadrupolar and toroidal moments in the $x-y$ plane, spin current with components along all axes and $\bm{\mathcal{P}}=0$. 
 \begin{figure}
\includegraphics[width=\columnwidth]{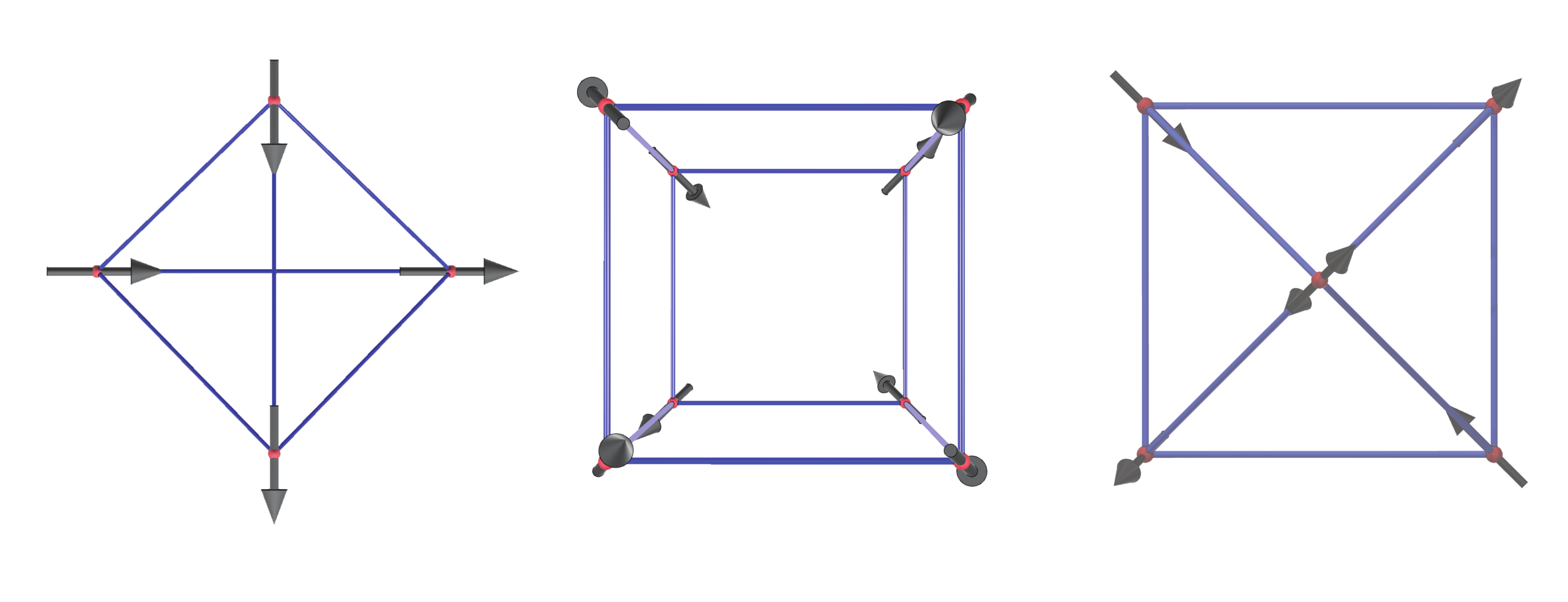}
\caption{x-y  projection (top view) of the minimum energy magnetic configuration states of dipoles at the sites of tetrahedral, cubic and octahedral clusters.}
\label{fig:f2}
\end{figure}

 \emph{Ground state sector.}  
The GS sector of a cluster $\mathcal{C}$, has hamiltonian $H_{GS}^{(\mathcal{C})}= \sum_{jk} \hat m_j A_{jk}^{(\mathcal{C})} \hat m_k$.  Coefficients,  $A_{j,k}^{(\mathcal{C})}$, are equal to the dipolar energy between dipoles $j$ and $k$ in the ground state of cluster $\mathcal{C}$, further $A^{(\mathcal{C})}$ is symmetric and has no diagonal elements. We gain insight into the symmetries of even polygons and polyhedra by solving the spectrum of the interaction matrix $A^{(\mathcal{C})}$. Indeed diagonalization of such a matrix for all clusters, yields eigenvalues with even degeneracies which do not correspond to the dimensions of the $irreps$ of  the corresponding point groups. 
We address this issue by building an effective Hamiltonian $\mathcal{\hat{H}}_{f}$ where matrix elements consist of link variables that represent Ising interactions among collinear dipoles. As an example, consider the hexagonal cluster in Fig.~\ref{fig:f1}. Exact diagonalization of the interaction matrix $A^{(h)}$ yields a spectrum with multiplicities $\{1,2,2,1\}$, where the GS has degeneracy 2.  The point group of the hexagon is the dihedral group \textit{622}($D_6$) with $h=12$ symmetry elements. Of the 12 symmetries, the magnetic configuration preserves two 3-fold rotations respect to the principal axis ($\hat{z}$), and two 2-fold rotations respect to axes perpendicular to $\hat{z}$.  Take now any pair of collinear dipoles (there are three of them in the hexagon),  say the $(p,q)$ pair, and associate an Ising variable (matrix element) $u_{p,q}$ according to the following rules: if $\hat{\bm{m}}_p\cdot \hat{\bm{m}}_q=1$, $u_{p,q}=i=-u_{q,p}$; if $\hat{\bm{m}}_p\cdot \hat{\bm{m}}_q=-1$, $u_{p,q}=1=u_{q,p}$ provided $p$ and $q$ point into the cluster and $u_{p,q}=-1=u_{q,p}$ provided $p$ and $q$ point out of the cluster, in all other cases $u_{p,q}=0$. Applying this procedure to the hexagon yields a $6\times 6$ hermitic matrix consisting of three $\sigma_y$ Pauli matrices along its diagonal, namely $\mathcal{\hat{H}}_{f}^{(h)}= (\hat{\sigma}_y,\hat{\sigma}_y,\hat{\sigma}_y)$. The spectrum of $\mathcal{\hat{H}}_{f}^{(h)}$ is $\{-1,-1,-1,1,1,1\}$, and therefore it preserves the degeneracies of the eigenvalues of $A^{(h)}$. $\mathcal{\hat{H}}_{f}$ reduces $\mathcal{H}_{GS}$ to its diagonal form and, more important, it  produces a representation easy to deal with symmetry wise. Applying the twelve symmetries of \textit{622} to the hexagon permute its sites.  We denote these point symmetry operations by $R$.  Some of these permutations will change the sense of Ising variables in $\mathcal{\hat{H}}_{f}^{(h)}$. However, spin current, ME moments and ME polarization remain unchanged under $R$. For those pairs of sites affected, we can apply the $\hat{\sigma}_z$  operator which flips the Ising variables and fix the problem. For instance, if the operation $R$ reverses $u_{1,2}$ connecting sites 1 and 2, we can fix it by combining $R$ (in the regular representation) with the matrix $\Lambda_{12}=(\hat{\sigma}_z ,\hat{1}_2,\hat{1}_2)$ ($\hat{1}_2$ the $2\times2$ identity matrix). The combined operation $\mathcal{R}=\Lambda_{12} R$ leaves $\mathcal{\hat{H}}_{f}^{(h)}$ invariant, as it does the equivalent combined symmetry $\mathcal{R}=\Lambda'_{12}R=\Lambda \Lambda_{12}R$ with $\Lambda=(-\hat{1}_2,-\hat{1}_2,-\hat{1}_2)$.  Further,  $\Lambda_{12}$ and $\Lambda$ are gauge transformations.  Applying this procedure to the symmetry operations of the point group \textit{622}, one finds that the six symmetries that alter Ising variables in the hexagonal cluster can be fixed by combining them with gauge transformations. Those $R$ that leave $u$ unchanged, can be combined with gauge transformation $\Lambda$ to yield its twin symmetry $\mathcal{R}=\Lambda R$.  Doing so for all symmetries of \textit{622} one finds that the number of symmetry elements of the hexagonal dipolar cluster has been doubled to 24 elements, it has three additional classes and three additional $irreps$ of dimension two each. Indeed it has become the double group \textit{622}, a result that makes sense in the light of the spin orbit interaction shown in Eq.~(\ref{eq:EnergyEven}). Table~\ref{table:2} shows the symmetry groups and $\mathcal{\hat{H}}_{f}$ for all clusters studied here. 
\begin{table}[bt]
\begin{tabular}{|c|c|c|c|c|c|}
\hline
$\mathcal{C}$ & $a$&$t$&$q_{\alpha, \beta}$& $\mathcal{J}_s$&$\mathcal{P}$\\
\hline
Odd Polygons 
	& 0
	& $t_x,t_y$ &$q_{x^2-y^2}$,$q_{x^2-z^2}$& $j_x,j_y,j_z$ &$p_x,p_y,p_z$
	\\ 
\hline
Triangle 
	& 0
	& $t_y$ &0& $j_x,j_y,j_z$ &$p_x,p_z$
	\\
\hline
Even Polygons
	& 0 
	& $0$ &0 &$j_z$ &$p_x,p_y$
	\\
\hline
Square
	& 0
	& $0$ &$q_{x^2-y^2},q_{x^2-z^2}$&$j_z$ &$p_x,p_y$
	\\	
\hline
Tetrahedron
	& 0
	& $t_y$ & 0 & $j_z$ &$p_x$
	\\
\hline
Cube
	& 0
	& $0$ &0& $j_x,j_y$ &$p_x,p_y$
	\\
\hline
Octahedron
	& yes
	& $t_x,t_y$ &$q_{x^2-z^2}$& $j_x,j_y,j_z$ &0
	\\
\hline
\end{tabular}
\caption{ME moments, spin current and polarization for polyhedral and polygonal clusters. Detailed expressions can be found in \cite{supp}.}
\label{table:1}
\end{table}

\begin{table}[bt]
\begin{tabular}{ |c|c|c|c| }
\hline
$\mathcal{C}$&$H_{f}^{(\mathcal{C})}$&$SG$&$\underline{Q}$\\
\hline
Square
	&$(\hat{\sigma}_x,-\hat{\sigma}_x)$  
	&\it$\bar{4}$2m &$
(Q_{11},-Q_{11})
$   
	\\
\hline
Hexagon
	&$(\hat{\sigma}_y,\hat{\sigma}_y,\hat{\sigma}_y)$  
	&\it622 &$(Q_{11},Q_{22})$  
	\\	
\hline
Tetrahedron
	& $(\hat{\sigma}_y,\hat{\sigma}_y)$ 
	&\it422 & $(Q_{11},Q_{11},Q_{22})$  
	\\
\hline
Cube
	&$(\hat{\sigma}_x,-\hat{\sigma}_x,\hat{\sigma}_x,-\hat{\sigma}_x)$  
	&\it$\bar{4}$2m  &$(Q_{11},-Q_{11},0)$   
	\\
\hline
Octahedron
	&  $(\hat{\sigma}_x,-\hat{\sigma}_x,\hat{\sigma}_x)$  
	&\it$\bar{4}$2m  & $(Q_{11},-Q_{11},0)$ 
	\\
\hline
\end{tabular}
\caption{Diagonal elements of $H_{f}^{(\mathcal{C})}$, cluster's double groups of symmetry (SG) (international notation) and diagonal elements of $\underline{Q}$ \cite{supp}.}
\label{table:2}
\end{table}
\emph{Double groups and $\underline{Q}$.} The program implemented on the hexagon, was applied to all even polygons and polyhedra examined in this paper. In polygons with $n=4s$ vertices, the symmetry group corresponds to the double group \textit{$\bar{n}$2m}, while in polygons with $n=2(2s+1)$ vertices it is the double group \textit{n22}. \textit{n22} and \textit{$\bar{n}$2m} differ in that the first is chiral while the second is not. This has an impact in the shape of $\underline{Q}$. Indeed, $\underline{Q}$ is a second rank axial tensor which connects a polar vector with an axial vector. For a system, whose symmetries are determined by a point group $G$, a polar vector $\bm{E}$ and an axial vector $\bm{M}$ transform according to $irreps$ $\Gamma_E$ and $\Gamma_M$ of $G$. The number of independent matrix elements of $\underline{Q}$ is the number of times that the scalar $irrep$, $\Gamma_1$ is contained in the decomposition of the direct product $\Gamma_E \otimes \Gamma_M$. For \textit{n22} point groups, the number of times that $\Gamma_1$ is contained is equal to 2, while for \textit{$\bar{n}$2m}, $\Gamma_1$ is contained once. Therefore in the first case the number of independent matrix elements of $\underline{Q}$ is two, while in the second case is one. Applying the generators of symmetries of the respective point groups to $\underline{Q}$, ($\underline{Q}$ transforms according to the rules of an axial vector) it is straightforward to determine the positions of those coefficients in each case. 
For the tetrahedron, cube and octahedron the symmetry groups are \textit{422}, \textit{$\bar{4}$2m}, \textit{$\bar{4}$2m} respectively. Table~\ref{table:2} shows the ME tensor in all cases \cite{supp}. 

 \emph{Conclusions.}  We have shown that magnetic dipoles at the sites of two and three dimensional clusters, some of them motif of crystallographic space groups,  are active for ME effect, carry spin current and in several cases manifest antisymmetric and symmetric ME moments. Using $\gamma$ as the relevant energy scale, we estimate that our systems achieve thermal stability for magnetic degrees of freedom on a scale of 10 [nm].  For polarization, in units of [$\gamma/L$],  and spin current, in units of [$\gamma/L^2$], we find that typical magnitudes are of the order of unity. We can estimate these values for magnetic nanoarrays: taking for instance permalloy nanoislands with $L=100 \times 10^{-9}$ [m] one estimate the spin currents to be of the order of $\bm{\mathcal{J}}_s\sim 4\times 10 ^{-25}$ [Joule m] and the polarizations about $\bm{\mathcal{P}}\sim 4 \times 10^{-32}$ [Joule $m^2$]. We found that the symmetries of the GS sector of these clusters are realized by double point groups, extensions of ordinary point groups that accommodate states with half-integer angular momentum, and consequently hold even dimensional representations. The dipolar hamiltonian in these systems exposes a spin-orbit coupling that manifests in a {Dzyaloshinskii-Moriya} interaction which explains the onset of double group symmetries. The origin of half-integer angular momentum associated to even dimensional $irreps$ can be explained in terms of  the spin current in these clusters. Indeed, in two dimensions a magnetic flux $\Phi$ can be defined from the spin current across the cluster. For the case of even polygons
 with circumradius $\rho=\frac{L}{2\sin(\pi/n)}$ 
 this flux becomes~: $\Phi= \frac{\rho}{m_0} \mathcal{J}_z =\frac{(n-1)\sin \left(\frac{ 2\pi }{n}\right)}{8\pi\sin \left(\frac{ \pi }{n}\right)}$, in units of $[\frac{\mu_0 m_0}{ L}]$. $\Phi$  is proportional to a magnetic charge $g$,  $2\pi\Phi=\rm g$,  which in the large $n$ limit, ${\rm g}\to \frac{(n-1)}{2}$. In even clusters $\rm g$  takes half-integer values  in units of $[\frac{\mu_0 m_0}{ L}]$ and it is responsible of a change of the net angular momentum of our clusters from integer to half-integer values \cite{mellado2015}.
 
 The detection of ME effect could be performed by Raman spectroscopy, which is extremely sensitive to changes in the electric polarization as demonstrated in \cite{cambre2015asymmetric}
or using similar optical probes, as has been shown in \cite{wei2017} for samples
under an increasing external magnetic field. Given their geometry and magnetic properties, we suggest experiments in Copper Keplerates \cite{palacios2016copper} or related compounds \cite{muller1999archimedean}.
 
\emph{Acknowledgments.}  This work was supported in part by Fondecyt under Grant No. 11121397 (PM), and by Fondecyt under Grant No. 1181382 (SR). P.M. acknowledges support from the Simons Foundation.

\end{document}